\ifx\extended\relax
\documentclass[11pt,a4size]{article}
\usepackage{authblk}
\usepackage[numbers]{natbib}
\usepackage{hyperref}
\usepackage{graphicx}
\usepackage{listings}
\usepackage{xcolor}
\usepackage{needspace}
\usepackage{rotating}
\usepackage{pdflscape}
\usepackage{afterpage}
\newcommand\blfootnote[1]{%
  \begingroup
  \renewcommand\thefootnote{}\footnote{#1}%
  \addtocounter{footnote}{-1}%
  \endgroup
}
\definecolor{dkgreen}{rgb}{0,0.6,0}
\definecolor{gray}{rgb}{0.5,0.5,0.5}
\definecolor{mauve}{rgb}{0.58,0,0.82}

\usepackage{url}

\definecolor{background}{HTML}{EEEEEE}

\else
\documentclass[sigconf]{acmart}
\fi

\newcommand{\kwlist}{Software engineering economics, software ecosystems,
open source software in business, Fortune Global 500,
SEC 10-K, SEC 20-F, EDGAR, dataset}

\ifx\extended\relax
\newcommand\extremark[1]{ (e.g. \emph{#1})}
\newcommand\extbegin[1]{\begin{#1}}
\newcommand\extend[1]{\end{#1}}
\newcommand\extitem{\item}
\newcommand\descitem[1]{\item[#1]}
\else
\newcommand\extremark[1]{}
\newcommand\extbegin[1]{}
\newcommand\extend[1]{}
\newcommand\extitem{}
\newcommand\descitem[1]{\texttt{#1},}
\fi

\AtBeginDocument{%
  \providecommand\BibTeX{{%
    \normalfont B\kern-0.5em{\scshape i\kern-0.25em b}\kern-0.8em\TeX}}}

\ifx\extended\undefined
\copyrightyear{2020}
\acmYear{2020}
\setcopyright{othergov}
\acmConference[MSR '20]{17th International Conference on Mining Software Repositories}{October 5--6, 2020}{Seoul, Republic of Korea}
\acmBooktitle{17th International Conference on Mining Software Repositories (MSR '20), October 5--6, 2020, Seoul, Republic of Korea}
\acmPrice{15.00}
\acmDOI{10.1145/3379597.3387495}
\acmISBN{978-1-4503-7517-7/20/05}
\fi

\begin{document}

\ifx\extended\relax
\title{A Dataset of Enterprise-Driven Open Source Software:\\
Extended Description}
\author{Diomidis Spinellis}
\author{Zoe Kotti}
\author{Konstantinos Kravvaritis}
\author{Georgios Theodorou}
\author{Panos Louridas}
\renewcommand\Affilfont{\itshape\small}
\affil{Athens University of Economics and Business}

\else

\title{A Dataset of Enterprise-Driven Open Source Software}

\author{Diomidis Spinellis}
\email{{dds,zoekotti,kravvaritisk,,louridas}@aueb.gr}
\orcid{0000-0003-4231-1897}
\author{Zoe Kotti}
\orcid{0000-0003-3816-9162}
\author{Konstantinos Kravvaritis}
\orcid{0000-0002-8889-0612}
\author{Georgios Theodorou}
\orcid{0000-0001-5413-2189}
\author{Panos Louridas}
\orcid{0000-0002-3971-4612}
\affiliation{%
 \institution{Athens University of Economics and Business}
%  \city{Athens}
%  \country{Greece}
}

\renewcommand{\shortauthors}{Spinellis et al.}
\fi

% Define commands for referring to tables
% Parenthetical reference to a database table
% Reference to a database table
\ifx\extended\relax
\input{lstrecordcount}
\else
\newcommand{\tblProjectCommitCommitterDomainCount}{\emph{project commit committer domain count}}
\newcommand{\ptblrecProjectCommitCommitterDomainCount}{}
\newcommand{\ptblProjectCommitCommitterDomainCount}{(Table \emph{project commit committer domain count})}
\newcommand{\tblEnterpriseProjectDetails}{\emph{enterprise project details}}
\newcommand{\ptblrecEnterpriseProjectDetails}{}
\newcommand{\ptblEnterpriseProjectDetails}{(Table \emph{enterprise project details})}
\newcommand{\tblSameDomainTopCommitters}{\emph{same domain top committers}}
\newcommand{\ptblrecSameDomainTopCommitters}{}
\newcommand{\ptblSameDomainTopCommitters}{(Table \emph{same domain top committers})}
\newcommand{\tblMergedDomainProjects}{\emph{merged domain projects}}
\newcommand{\ptblrecMergedDomainProjects}{}
\newcommand{\ptblMergedDomainProjects}{(Table \emph{merged domain projects})}
\newcommand{\tblForksClonesNoise}{\emph{forks clones noise}}
\newcommand{\ptblrecForksClonesNoise}{}
\newcommand{\ptblForksClonesNoise}{(Table \emph{forks clones noise})}
\newcommand{\tblProjectAuthorCount}{\emph{project author count}}
\newcommand{\ptblrecProjectAuthorCount}{}
\newcommand{\ptblProjectAuthorCount}{(Table \emph{project author count})}
\newcommand{\tblValidEnterpriseDomains}{\emph{valid enterprise domains}}
\newcommand{\ptblrecValidEnterpriseDomains}{}
\newcommand{\ptblValidEnterpriseDomains}{(Table \emph{valid enterprise domains})}
\newcommand{\tblSizeMetrics}{\emph{size metrics}}
\newcommand{\ptblrecSizeMetrics}{}
\newcommand{\ptblSizeMetrics}{(Table \emph{size metrics})}
\newcommand{\tblProbableCompanyDomainCommitters}{\emph{probable company domain committers}}
\newcommand{\ptblrecProbableCompanyDomainCommitters}{}
\newcommand{\ptblProbableCompanyDomainCommitters}{(Table \emph{probable company domain committers})}
\newcommand{\tblMultiCommitterProbableCompanyProjects}{\emph{multi committer probable company projects}}
\newcommand{\ptblrecMultiCommitterProbableCompanyProjects}{}
\newcommand{\ptblMultiCommitterProbableCompanyProjects}{(Table \emph{multi committer probable company projects})}
\newcommand{\tblDomainBlacklist}{\emph{domain blacklist}}
\newcommand{\ptblrecDomainBlacklist}{}
\newcommand{\ptblDomainBlacklist}{(Table \emph{domain blacklist})}
\newcommand{\tblUserDomain}{\emph{user domain}}
\newcommand{\ptblrecUserDomain}{}
\newcommand{\ptblUserDomain}{(Table \emph{user domain})}
\newcommand{\tblValidEnterpriseDomainCommitters}{\emph{valid enterprise domain committers}}
\newcommand{\ptblrecValidEnterpriseDomainCommitters}{}
\newcommand{\ptblValidEnterpriseDomainCommitters}{(Table \emph{valid enterprise domain committers})}
\newcommand{\tblProjectCommitCommitterDomain}{\emph{project commit committer domain}}
\newcommand{\ptblrecProjectCommitCommitterDomain}{}
\newcommand{\ptblProjectCommitCommitterDomain}{(Table \emph{project commit committer domain})}
\newcommand{\tblForksClonesNoiseNames}{\emph{forks clones noise names}}
\newcommand{\ptblrecForksClonesNoiseNames}{}
\newcommand{\ptblForksClonesNoiseNames}{(Table \emph{forks clones noise names})}
\newcommand{\tblCandidateProjects}{\emph{candidate projects}}
\newcommand{\ptblrecCandidateProjects}{}
\newcommand{\ptblCandidateProjects}{(Table \emph{candidate projects})}
\newcommand{\tblProjectCommitAuthorDomainCount}{\emph{project commit author domain count}}
\newcommand{\ptblrecProjectCommitAuthorDomainCount}{}
\newcommand{\ptblProjectCommitAuthorDomainCount}{(Table \emph{project commit author domain count})}
\newcommand{\tblProbableCompanyProjects}{\emph{probable company projects}}
\newcommand{\ptblrecProbableCompanyProjects}{}
\newcommand{\ptblProbableCompanyProjects}{(Table \emph{probable company projects})}
\newcommand{\tblSecTenKDomains}{\emph{sec 10 K domains}}
\newcommand{\ptblrecSecTenKDomains}{}
\newcommand{\ptblSecTenKDomains}{(Table \emph{sec 10 K domains})}
\newcommand{\tblProbableCompanyDomains}{\emph{probable company domains}}
\newcommand{\ptblrecProbableCompanyDomains}{}
\newcommand{\ptblProbableCompanyDomains}{(Table \emph{probable company domains})}
\newcommand{\tblSecTwentyFDomains}{\emph{sec 20 F domains}}
\newcommand{\ptblrecSecTwentyFDomains}{}
\newcommand{\ptblSecTwentyFDomains}{(Table \emph{sec 20 F domains})}
\newcommand{\tblValidEnterpriseProjects}{\emph{valid enterprise projects}}
\newcommand{\ptblrecValidEnterpriseProjects}{}
\newcommand{\ptblValidEnterpriseProjects}{(Table \emph{valid enterprise projects})}
\newcommand{\tblSharedDomains}{\emph{shared domains}}
\newcommand{\ptblrecSharedDomains}{}
\newcommand{\ptblSharedDomains}{(Table \emph{shared domains})}
\newcommand{\tblProjectCommitterDomains}{\emph{project committer domains}}
\newcommand{\ptblrecProjectCommitterDomains}{}
\newcommand{\ptblProjectCommitterDomains}{(Table \emph{project committer domains})}
\newcommand{\tblCohortProjects}{\emph{cohort projects}}
\newcommand{\ptblrecCohortProjects}{}
\newcommand{\ptblCohortProjects}{(Table \emph{cohort projects})}
\newcommand{\tblOrgDomains}{\emph{org domains}}
\newcommand{\ptblrecOrgDomains}{}
\newcommand{\ptblOrgDomains}{(Table \emph{org domains})}
\newcommand{\tblAllCompanyDomains}{\emph{all company domains}}
\newcommand{\ptblrecAllCompanyDomains}{}
\newcommand{\ptblAllCompanyDomains}{(Table \emph{all company domains})}
\newcommand{\tblPullRequestNumber}{\emph{pull request number}}
\newcommand{\ptblrecPullRequestNumber}{}
\newcommand{\ptblPullRequestNumber}{(Table \emph{pull request number})}
\newcommand{\tblFortuneGlobalFiveHundred}{\emph{fortune global 500}}
\newcommand{\ptblrecFortuneGlobalFiveHundred}{}
\newcommand{\ptblFortuneGlobalFiveHundred}{(Table \emph{fortune global 500})}
\newcommand{\tblProbableCompanyUsers}{\emph{probable company users}}
\newcommand{\ptblrecProbableCompanyUsers}{}
\newcommand{\ptblProbableCompanyUsers}{(Table \emph{probable company users})}
\newcommand{\tblProjectCommitDetails}{\emph{project commit details}}
\newcommand{\ptblrecProjectCommitDetails}{}
\newcommand{\ptblProjectCommitDetails}{(Table \emph{project commit details})}
\newcommand{\tblProjectStars}{\emph{project stars}}
\newcommand{\ptblrecProjectStars}{}
\newcommand{\ptblProjectStars}{(Table \emph{project stars})}
\newcommand{\tblMultiCommitterValidEnterpriseProjects}{\emph{multi committer valid enterprise projects}}
\newcommand{\ptblrecMultiCommitterValidEnterpriseProjects}{}
\newcommand{\ptblMultiCommitterValidEnterpriseProjects}{(Table \emph{multi committer valid enterprise projects})}
\newcommand{\tblCompanyTlds}{\emph{company tlds}}
\newcommand{\ptblrecCompanyTlds}{}
\newcommand{\ptblCompanyTlds}{(Table \emph{company tlds})}
\newcommand{\tblAboveAverageProjects}{\emph{above average projects}}
\newcommand{\ptblrecAboveAverageProjects}{}
\newcommand{\ptblAboveAverageProjects}{(Table \emph{above average projects})}
\newcommand{\tblProjectCommitterCount}{\emph{project committer count}}
\newcommand{\ptblrecProjectCommitterCount}{}
\newcommand{\ptblProjectCommitterCount}{(Table \emph{project committer count})}
\newcommand{\tblValidEnterpriseUsers}{\emph{valid enterprise users}}
\newcommand{\ptblrecValidEnterpriseUsers}{}
\newcommand{\ptblValidEnterpriseUsers}{(Table \emph{valid enterprise users})}
\newcommand{\tblDomains}{\emph{domains}}
\newcommand{\ptblrecDomains}{}
\newcommand{\ptblDomains}{(Table \emph{domains})}
\newcommand{\tblProjectSizeMetrics}{\emph{project size metrics}}
\newcommand{\ptblrecProjectSizeMetrics}{}
\newcommand{\ptblProjectSizeMetrics}{(Table \emph{project size metrics})}
\newcommand{\tblProjectCommitCount}{\emph{project commit count}}
\newcommand{\ptblrecProjectCommitCount}{}
\newcommand{\ptblProjectCommitCount}{(Table \emph{project commit count})}
\newcommand{\tblLicenses}{\emph{licenses}}
\newcommand{\ptblrecLicenses}{}
\newcommand{\ptblLicenses}{(Table \emph{licenses})}
\newcommand{\tblNoTable}{\emph{no table}}
\newcommand{\ptblrecNoTable}{}
\newcommand{\ptblNoTable}{(Table \emph{no table})}
\newcommand{\tblDeduplicatedProjects}{\emph{deduplicated projects}}
\newcommand{\ptblrecDeduplicatedProjects}{}
\newcommand{\ptblDeduplicatedProjects}{(Table \emph{deduplicated projects})}
\newcommand{\tblProjectCommitAuthorDomain}{\emph{project commit author domain}}
\newcommand{\ptblrecProjectCommitAuthorDomain}{}
\newcommand{\ptblProjectCommitAuthorDomain}{(Table \emph{project commit author domain})}
\newcommand{\tblDistinctCompanyDomains}{\emph{distinct company domains}}
\newcommand{\ptblrecDistinctCompanyDomains}{}
\newcommand{\ptblDistinctCompanyDomains}{(Table \emph{distinct company domains})}
\newcommand{\tblProjectAuthorDomains}{\emph{project author domains}}
\newcommand{\ptblrecProjectAuthorDomains}{}
\newcommand{\ptblProjectAuthorDomains}{(Table \emph{project author domains})}
\newcommand{\tblCommitRange}{\emph{commit range}}
\newcommand{\ptblrecCommitRange}{}
\newcommand{\ptblCommitRange}{(Table \emph{commit range})}
\newcommand{\tblProjectCommitterDomainRank}{\emph{project committer domain rank}}
\newcommand{\ptblrecProjectCommitterDomainRank}{}
\newcommand{\ptblProjectCommitterDomainRank}{(Table \emph{project committer domain rank})}
\newcommand{\tblMergedProjects}{\emph{merged projects}}
\newcommand{\ptblrecMergedProjects}{}
\newcommand{\ptblMergedProjects}{(Table \emph{merged projects})}

\fi

\ifx\extended\relax
\maketitle
\else


\fi

\begin{abstract}
We present a dataset of open source software
developed mainly by enterprises rather than volunteers.
This can be used to address known generalizability concerns,
and, also, to perform research on open source business software development.
Based on the premise that an enterprise's employees are likely
to contribute to a project developed by their
organization using the email account provided by it,
we mine domain names associated with enterprises from
open data sources as well as through white- and blacklisting,
and use them through three heuristics
to identify 17\,264 enterprise GitHub projects.
We provide these as a dataset detailing their provenance and properties.
A manual evaluation of a dataset sample
shows an identification accuracy of 89\%.
Through an exploratory data analysis we found that
projects are staffed by a plurality of enterprise insiders,
who appear to be pulling more than their weight, and that in a small
percentage of relatively large projects development happens exclusively
through enterprise insiders.
\end{abstract}

\ifx\extended\relax


\blfootnote{This is a technical note expanding reference~\cite{SKKT20p},
which should be cited in preference to this text.}

\else

\maketitle

\fi

\section{Introduction} % {{{1
\label{sec:intro}
Despite the size and wealth of software product and process
data available on GitHub,
their use in software engineering research can be
problematic~\cite{KGBS16,CIC16},
raising issues regarding the generalizability of the corresponding
findings~\cite{WKP10}.
In particular, the open source nature of accessible GitHub
repositories means that projects developed by volunteers through
open source software development processes~\cite{FF02b,SFFH06}
are overrepresented,
biasing results, especially those related to
software architecture or communication and organization structures,
through the application of Conway's Law~\cite{Con68,HG99}.
In addition, many researchers are investigating differences
between open source and proprietary software products and
processes~\cite{PSE04,MRB06,Spi08b,BB09}.

Here we present a dataset of open source software
developed mainly by enterprises rather than volunteers.
This can be used to
address the identified generalizability concerns
and, also, to perform research on the differences between
volunteer and business software development.
One might think that open source software development by enterprises
is a niche phenomenon.
As others have identified~\cite{RKCC18} and also as is evident from
our dataset,
this is far from true.
A series of queries on GitHub PushEvents published during 2017
found that companies such as Microsoft and Google had hundreds
of employees contributing to open source projects~\cite{Hof17}.

The goal of the dataset's construction is to create a set
of GitHub projects that are most probably developed by
an enterprise.
For the purposes of this work, we define as an enterprise project,
one that is likely to be mainly developed by financially compensated employees,
working full time under an organization's management.
This definition excludes volunteer efforts such as
Linux,
KDE projects,
VLC, and
GIMP
(even though some companies pay their employees to contribute to them),
but includes for-profit company and funded public-sector
organization projects that accept volunteer contributions,
such as
Google's Trillian,
Apple's Swift,
CERN's ALICE, and
Microsoft's Typescript.
Our aim is to minimize the number of false positives in the dataset,
but we are not interested in the number of false negatives.
We do not aspire to create a comprehensive dataset of enterprise
projects, but one that contains a number sufficient to conduct
generalizable empirical studies.

\ifx\extended\relax
In the following sections we present
how we collected the data (Section~\ref{sec:method})
and evaluated them (Section~\ref{sec:evaluation}),
the data schema and availability (Section~\ref{sec:overview}),
as well as indicative findings (Section~\ref{sec:findings}),
related work (Section~\ref{sec:related}), and
ideas for research and improvements (Section~\ref{sec:ideas}).
\fi

\ifx\extended\relax
\section{Dataset Construction} % {{{1
\else
\section{Construction and Evaluation} % {{{1
\label{sec:evaluation}
\fi
\label{sec:method}
\ifx\extended\relax
\afterpage{% https://tex.stackexchange.com/a/239366/10140
\begin{landscape} % https://tex.stackexchange.com/a/159992/10140
\begin{figure}[p]
\includegraphics[width=\columnwidth]{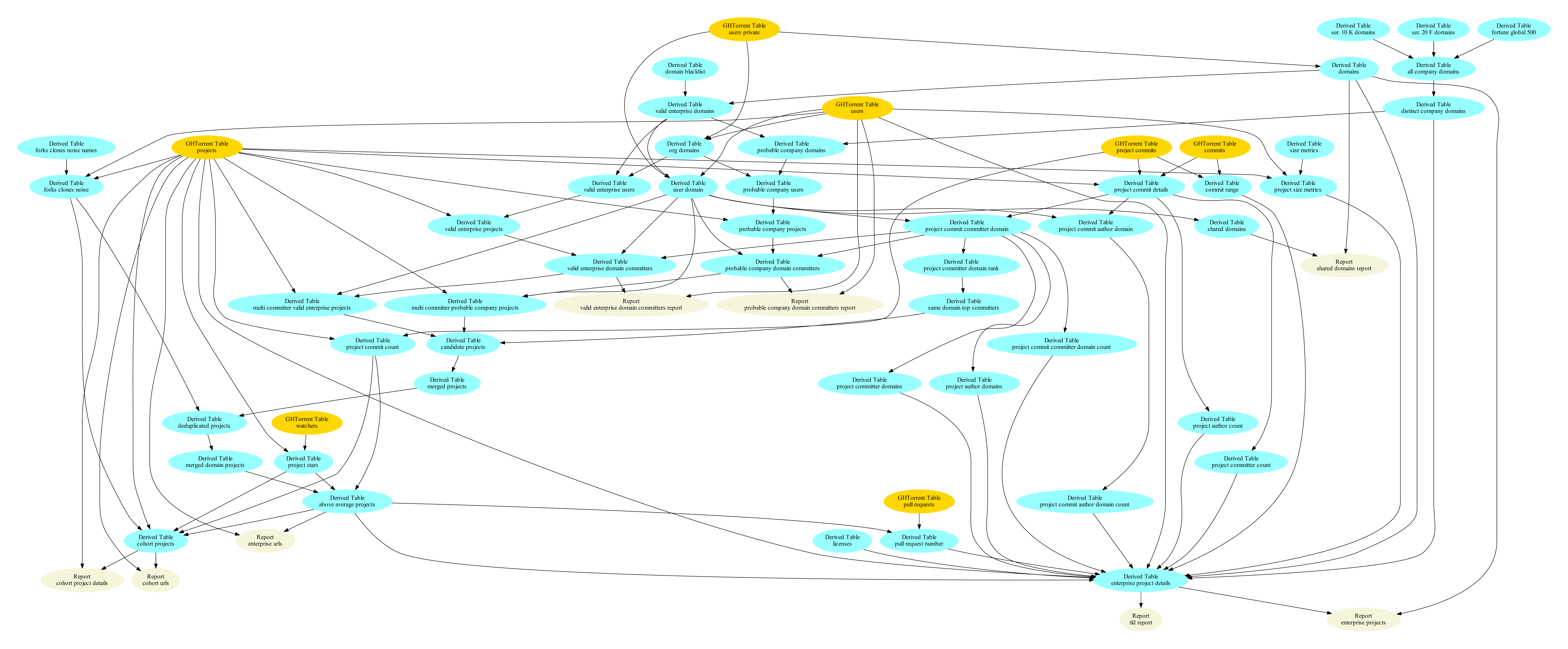}
\caption{Overview of the dataset creation process}
\label{fig:dataflow}
\end{figure}
\end{landscape}
}
\fi
An overview of the dataset's construction process is depicted in
\ifx\extended\relax
Figure~\ref{fig:dataflow}.
\else
an extended version of this paper~\cite{SKKT20b}.
\fi
The projects were selected from GitHub by analyzing
the GHTorrent~\cite{GS12,Gou13} dataset (release 2019-06-01)
by means of the {\em simple-rolap} relational online analytical processing
and {\em rdbunit} relational unit testing frameworks~\cite{GS17}.
Following published
\ifx\extended\relax
recommendations~\cite{IHG12},
\else
recommendations
\fi
the code and primary data associated with this endeavor
are openly available online.\footnote{\url{https://doi.org/10.5281/zenodo.3742973}}

The basic premise for constructing the dataset is that an enterprise's
employees are likely to contribute to a project developed by their
organization using the email account provided by it.
Furthermore, it is unlikely that pure volunteer projects will
have contributors using emails from a single enterprise-related
domain address.
Based on this premise, we identified projects where a large number of
commits were contributed through accounts linked to the same enterprise
email domain address.
To increase the dataset's quality we then removed project clones~\cite{SKM20},
and only retained projects having more than the identified dataset's average
stars (14) and commits (29).
Finally, we created one table with diverse details regarding each
selected project and one with details regarding each associated
enterprise domain.
The following paragraphs detail each step,
starting from the creation of two tables: {\em valid enterprise domains}
and {\em probable company domains}.\footnote{In the interest of readability,
this text replaces the underscores in the table names with spaces.}

\paragraph{Valid enterprise domains} This table \ptblrecValidEnterpriseDomains{}
was created by filtering
all email domains found in the users' email table \ptblDomains{}.
We did this by examining frequently occurring email domains,
and creating rules to retain only those associated with enterprise development.
Specifically, we removed from the set of domains a blacklist
\ptblDomainBlacklist{} containing those associated with:
\extbegin{itemize}
\extitem email providers\extremark{gmail.com, qq.com, outlook.com, yandex.ru}{};
\extitem top and second level organization domains\extremark{.org, .org.nz, .or.at}{},
    and thereby the many associated with volunteer open
    source organizations\extremark{apache.org, openssl.org}{};
\extitem open source hubs\extremark{sourceforge.net}{};
\extitem top and second level educational domains\extremark{.edu, .edu.au, .ac.uk}{}
    and, explicitly, the domains of more than 20 hand-picked
    universities\extremark{eurecom.fr, tu-dortmund.de}{};
\extitem individuals\extremark{schildbach.de}{}.
\extend{itemize}
We did not remove government organizations\extremark{lexingtonky.gov}{} and
research centers\extremark{cern.ch}{} as these mainly operate as enterprises
with professional developers.
When in doubt, we looked up company emails in the RocketReach provider
of company email format details.

\paragraph{Probable company domains} This table \ptblrecProbableCompanyDomains{}
was created by identifying
domains that are likely to belong to companies from publicly available
data and domain heuristics.
We obtained the domains associated with large companies in two ways.
First, we screen-scrapped, downloaded, and filtered the data associated with
   the Fortune Global 500 companies:
   the largest corporations across the globe measured by revenue
   \ptblFortuneGlobalFiveHundred{}.
Second, we obtained the US Securities and Exchange Commission (SEC) yearly company
   filings that are made in machine readable form
   \ifx\extended\relax
   (in the XBRL---eXtensible Business Reporting Language---an application of XML)
   \fi
   and extracted from them the company domains.
   Specifically, we obtained from EDGAR---the SEC's Electronic Data Gathering,
   Analysis, and Retrieval system---the XBRL files associated with two forms,
   namely
   a) Form 10-K, that gives a comprehensive summary of a company's financial
     performance \ptblSecTenKDomains{}, and
   b) Form 20-F, that provides an annual report filing for foreign private issuers---non-U.S.
     and non-Canadian companies that have securities trading
     in the U.S. \ptblSecTwentyFDomains{}.

   We then extracted the internet domain (e.g. {\em intel.com}) associated
   with each company from the XBRL files.
   \ifx\extended\relax
   Although the SEC provides guidance for using a corporate web site to
   disseminate public information,\footnote{\url{https://www.sec.gov/rules/interp/2008/34-58288.pdf}}
   it does not appear to collect these sites in a structured manner, within
   e.g. the XBRL files.
   \fi
   We obtained the company domains by looking at the XML name space
   used in the files, which in most cases contains the company's domain.
We combined the three sources into the Table \tblDistinctCompanyDomains{}
and complemented it with the Table \tblValidEnterpriseDomains{}
filtered to include only records associated with top and second level
commercial domains such as {\em .com}, {\em .co.uk, .com.au}.

\paragraph{From enterprise organizations to their projects}
As a next step we combined the two tables with another
listing domains registered for GitHub organizations \ptblOrgDomains{},
to get tables with user domains linked to GitHub organizations---Tables
\tblValidEnterpriseUsers{} and
\tblProbableCompanyUsers{}.
The intuition here is that many companies developing software on GitHub
will have configured a company organization under their domain name.
Combining the two tables with the GHTorrent \emph{Projects} table yielded
the corresponding projects hosted under a GitHub organization:
\tblValidEnterpriseProjects{} and
\tblProbableCompanyProjects{}.

These two tables were then linked with
a table of each user's email domain \ptblUserDomain{} and
one identifying each commit's committer \ptblProjectCommitCommitterDomain{},
giving the number of committers in each project associated with the
corresponding organization: \tblValidEnterpriseDomainCommitters{} and
\tblProbableCompanyDomainCommitters{}.
This stage ended by selecting projects from organizations having
a minimum number of committers appearing on GitHub with an email
associated with the organization's domain
giving the tables \tblMultiCommitterValidEnterpriseProjects{} and
\tblMultiCommitterProbableCompanyProjects{}.
The employed floor values (ten and five correspondingly) were selected
to exclude projects associated with individuals operating under
a personal but commercial-looking domain (e.g. {\em johnsmith.com}).

\paragraph{Enterprise-dominated projects}
To cover enterprises that may not have GitHub organizations
registered with emails under their domain,
we also established in each project a rank of committers
with valid enterprise email addresses
according to their number of commits \ptblProjectCommitterDomainRank{},
and obtained those projects having committers from the
same organizations as the topmost three \ptblSameDomainTopCommitters{}.

\paragraph{Final filtering and reporting}
For the three types of possible enterprise projects we then
formed their union \ptblCandidateProjects{},
combined their metrics \ptblMergedProjects{},
removed duplicate projects \ptblDeduplicatedProjects{},
combined records referring to the same project \ptblMergedDomainProjects{}, and
joined them with the number of their commits \ptblProjectCommitCount{}
 and their stars \ptblProjectStars{}, to select those with above
 average such metrics \ptblAboveAverageProjects{}.
For each one of the shortlisted projects, we
{\tt git-cloned} from GitHub the project's repository and
calculated its basic size metrics in terms of files and text lines
\ptblSizeMetrics{}.
(Due to churn from the date the GHTorrent dataset was published,
not all repositories could be retrieved for measuring project size.)

Finally, to provide context for each project,
we combined this table with each project's
earliest and most recent commit \ptblCommitRange{},
number of commits \ptblProjectCommitCommitterDomainCount{} and
committers \ptblProjectCommitterDomains{} for each committer domain,
number of commits \ptblProjectCommitAuthorDomainCount{} and
committers \ptblProjectAuthorDomains{} for each author domain,
total number of
  committers \ptblProjectCommitterCount{} and
  authors \ptblProjectAuthorCount{},
size metrics \ptblProjectSizeMetrics{},
project license as provided by the GitHub API \ptblLicenses{},
as well as details about the derivation of the corresponding domain.
This process created the table \tblEnterpriseProjectDetails{} and the
corresponding report \emph{enterprise projects}.

\ifx\extended\relax
\section{Evaluation} % {{{1
\label{sec:evaluation}
\fi
We manually evaluated a random sample of an earlier version of this
dataset,\footnote{\url{https://doi.org/10.5281/zenodo.3653878}
and \url{https://doi.org/10.5281/zenodo.3653888}.
This was updated following the peer review suggestions, and
differs by 64 projects (0.37\%---26 removed, 38 added)
from the currently supplied one.}
following the systematic review guidelines
by Brereton et al.~\cite{BKBT07}.
The sample size was calculated at around 378
using Cochran's sample size and correction formulas~\cite{Coc77}
(95\% confidence, 5\% precision).
To keep the raters alert
we complemented the sample with 22 GitHub projects
randomly selected from a set of projects with similar
quality characteristics that were part of the
dataset~\ptblCohortProjects{}.
The third and fourth authors were instructed to individually label
the 400 projects as enterprise or not based on the definition in
Section~\ref{sec:intro}.
To improve the labeling's reliability the two raters did
not know the employed heuristics, and were also asked to complete
the main reason the project was open source and
write a few words to support their decision.
% https://docs.google.com/spreadsheets/d/1F9VEURE0zmEA6HY3QjJUQHDGFgdX0YfODV-06aX3fII/edit?usp=sharing
Their ratings led to 78\% inter-rater agreement and 29\% reliability
using Cohen's kappa statistic. The second author then resolved the
conflicts by majority vote; after excluding the 22 irrelevant
projects, 89\% of the 378 projects were finally identified as
enterprise. 
We used the bootstrap method~\cite{Efr79} with 1000 iterations to
establish a confidence interval (CI) for the percentage of enterprise
projects in our sample; the 95\% CI was calculated at [87--93]\%.
% python ../src/sample_bootstrap.py
To generalize,
15\,354 (CI: 15\,009--16\,044) projects of our dataset are expected
to be truly enterprise-developed.

Regarding the dataset's external validity,
note that although our evaluation addresses the dataset's precision,
our method was not targeting a high recall and this was also not
evaluated.
Consequently, the dataset can be used to address empirical research
generalizability concerns we identified in the introduction mainly
by providing a set of enterprise-developed projects to be used
in work employing stratified sampling,
in cohort studies,
or in case studies.
Furthermore, the number of committers floor we employed in our selection
means that the dataset excludes organizations that are small or
have a tiny number of their employees committing on GitHub.
Finally, the selection of above average projects in terms of stars and
commits means that the dataset does not include stillborn or unpopular
projects.

\section{Dataset Overview} % {{{1
\label{sec:overview}
% select * from indoss.enterprise_project_details limit 1;
The dataset\footnote{\url{https://doi.org/10.5281/zenodo.3742962}}
is provided as a 17\,264 record
tab-separated file with the following 29 fields:
\extbegin{description}
\descitem{url} the project's GitHub URL;
\descitem{project\_\-id} the project's GHTorrent identifier;
% awk -F$'\t' '$3 == "t" {c++} END {print c}' ../src/reports/enterprise_projects.txt
\descitem{sdtc} true if selected using the same domain top committers heuristic (9\,016 records);
\descitem{mcpc} true if selected using the multiple committers from a valid enterprise heuristic (8\,314 records);
\descitem{mcve} true if selected using the multiple committers from a probable company heuristic (8\,015 records);
\descitem{star\_\-number} number of GitHub watchers;
\descitem{commit\_\-count} number of commits;
\descitem{files} number of files in current main branch;
\descitem{lines} corresponding number of lines in text files;
\descitem{pull\_\-requests} number of pull requests;
\descitem{github\_repo\_creation} time stamp of the GitHub repository creation;
\descitem{earliest\_\-commit} time stamp of the earliest commit;
\descitem{most\_\-recent\_\-commit} time stamp of the most recent commit;
\descitem{committer\_\-count} number of different committers;
\descitem{author\_\-count} number of different authors;
\descitem{dominant\_\-domain} the project's dominant email domain;
\descitem{dominant\_\-domain\_\-committer\_\-commits} number of commits made by committers whose email matches the project's dominant domain;
\descitem{dominant\_\-domain\_\-author\_\-commits} corresponding number for commit authors;
\descitem{dominant\_\-domain\_\-committers} number of committers whose email matches the project's dominant domain;
\descitem{dominant\_\-domain\_\-authors} corresponding number for commit authors;
\descitem{cik} SEC's EDGAR ``central index key'';
\descitem{fg500} true if this is a Fortune Global 500 company (2\,233 records);
\descitem{sec10k} true if the company files SEC 10-K forms (4\,180 records);
\descitem{sec20f} true if the company files SEC 20-F forms (429 records);
\descitem{project\_\-name} GitHub project name;
\descitem{owner\_\-login} GitHub project's owner login;
\descitem{company\_\-name} company name as derived from the SEC and Fortune 500 data;
\descitem{owner\_\-company} GitHub project's owner company name;
\descitem{license} SPDX license identifier.
\extend{description}
% $ wc -l reports/cohort_project_details
\ifx\extended\relax
An additional file provides the full set of 311\,223 cohort projects
(not part of the enterprise dataset), selected as described
in Section~\ref{sec:evaluation}, with the following four fields:
\extbegin{description}
\descitem{url} the project's GitHub URL;
\descitem{project\_id} the project's GHTorrent identifier;
\descitem{stars} number of GitHub watchers;
\descitem{commit\_count} number of commits.
\extend{description}
\fi

\ifx\extended\relax
\section{Typology of Enterprise OSS} % {{{1
\label{sec:findings}
We performed a preliminary analysis of the details we collected to obtain
a picture of how enterprise software is developed.
\fi
Overall, we see that projects are staffed by a plurality of enterprise insiders,
who appear to be pulling more than their weight.
Regarding the distribution of contributors,
across all identified projects in the dataset we found that
% select sum(dominant_domain_authors) / (select sum(author_count) from indoss.enterprise_project_details) from indoss.enterprise_project_details;
33\% of the authors and
% select sum(dominant_domain_committers) / (select sum(committer_count) from indoss.enterprise_project_details) from indoss.enterprise_project_details;
24\% of the committers
are associated with the project's dominant domain.
Similarly, regarding the distribution of work,
% select sum(dominant_domain_author_commits) / (select sum(commit_count) from indoss.enterprise_project_details) from indoss.enterprise_project_details;
% 0.44721603683743728767
45\% of the commits are made by the enterprise's authors,
and
% select sum(dominant_domain_committer_commits) / (select sum(commit_count) from indoss.enterprise_project_details) from indoss.enterprise_project_details;
% 0.41062220376412511936
41\% of the commits are made by the corresponding committers.

The ten most popular out of the 110
% awk -F$'\t' '{print $16}' enterprise_projects.txt  | sed 's/.*\.//' | sort | uniq | wc -l
top level domains associated with projects are:
% awk -F$'\t' '{print $16}' enterprise_projects.txt  | sed 's/.*\.//' | sort | uniq -c | sort -rn | head -10 | awk '{print "\\emph{" $2 "} (" $1 "),"}'
\emph{com} (13\,494 projects),
\emph{io} (763),
\emph{de} (383),
\emph{gov} (339),
\emph{net} (256),
\emph{ru} (142),
\emph{fr} (134),
\emph{cn} (120),
\emph{br} (118), and
\emph{uk} (111).
% awk -F$'\t' '{print $26}' enterprise_projects.txt  | sort | uniq | wc -l
Similarly, out of 5\,097 owners,
those associated with the highest number of GitHub projects are:
% awk -F$'\t' '{print $26}' enterprise_projects.txt  | sort | uniq -c | sort -rn | head -10 | awk '{print "\\emph{" $2 "} (" $1 "),"}'
\emph{Microsoft} (855 projects),
\emph{Azure} (328),
\emph{google} (123),
\emph{twitter} (93),
\emph{18F} (90),
\emph{udacity} (82),
\emph{SAP} (79),
\emph{Netflix} (79),
\emph{hashicorp} (77), and
\emph{GoogleCloudPlatform} (77).

In very few projects does development appear to be exclusively controlled
by the enterprise:
% select count(*), count(*) * 100. / (select count(*) from indoss.enterprise_project_details) from indoss.enterprise_project_details where commit_count = dominant_domain_committer_commits;
we found 90 projects (0.5\%) where all commits came from an enterprise
committer and
% select count(*), count(*) * 100. / (select count(*) from indoss.enterprise_project_details) from indoss.enterprise_project_details where commit_count = dominant_domain_author_commits;
220 projects (1.3\%) where all commits came from an enterprise author.
% select avg(lines)  from indoss.enterprise_project_details where commit_count = dominant_domain_author_commits;                                      -[ RECORD 1 ]------------
% avg | 452958.860962566845
We were expecting these projects to be small, but in fact
they sport an average line count of 453k for projects with exclusively
% select avg(lines)  from indoss.enterprise_project_details where commit_count = dominant_domain_committer_commits;
% vg | 975673.194805194805
enterprise authors and 976k for projects with exclusively enterprise
committers.
Considerable development seems to happen through pull requests, with
95\% of the projects having pull requests associated
% select (select sum(pull_requests) from indoss.enterprise_project_details) / (select count(*) from indoss.enterprise_project_details);                      ?column?
% 161.4685253883607698
with them, with an average of 161 pull requests per project.

% select license, count(*) from indoss.licenses group by license order by count desc;
In total,
according to their
\ifx\extended\relax
SPDX~\cite{GKKY14,KKT17}
\else
SPDX
\fi
identifiers,
% awk -F$'\t' '{print $29}' enterprise_projects.txt  | sort | uniq -c |wc -l
the projects are licensed using 29 different open source licenses.
The two most common licenses used are the
% awk -F$'\t' '{print $29}' enterprise_projects.txt  | sort | uniq -c | sort -rn | head
MIT (4\,340 projects) and
Apache 2.0 (3\,761 projects), with the GPL version 2 or 3 license used only by
780 projects.
This finding indicates that few enterprise open source projects seem to follow
a business model based on relicensing GPL code for proprietary development.
% empty
Surprisingly, for 3\,535 projects no license was found,
% NOASSERTION
and for 3\,374 projects the license did not match one with an SPDX identifier.

\begin{table}
\caption{\label{tab:metrics}Enterprise (E) and Reaper (R) Dataset Metrics}
\centering
\begin{tabular}{l rrrrrrrr}
\hline
% select min(star_number), min(commit_count), min(pull_requests), min(author_count), min(committer_count) from indoss.enterprise_project_details;
% select min(star_number), min(commit_count), min(pull_requests), min(author_count), min(committer_count) from reaper.reaper_project_details;
% select max(star_number), max(commit_count), max(pull_requests), max(author_count), max(committer_count) from indoss.enterprise_project_details;
% select max(star_number), max(commit_count), max(pull_requests), max(author_count), max(committer_count) from reaper.reaper_project_details;
% select avg(star_number), avg(commit_count), avg(pull_requests), avg(author_count), avg(committer_count) from indoss.enterprise_project_details;
% select avg(star_number), avg(commit_count), avg(pull_requests), avg(author_count), avg(committer_count) from reaper.reaper_project_details;
% select stddev(star_number), stddev(commit_count), stddev(pull_requests), stddev(author_count), stddev(committer_count) from indoss.enterprise_project_details;
% select stddev(star_number), stddev(commit_count), stddev(pull_requests), stddev(author_count), stddev(committer_count) from reaper.reaper_project_details;
	& \multicolumn{2}{c}{Min}	& \multicolumn{2}{c}{Max (k)}	& \multicolumn{2}{c}{Avg}	& \multicolumn{2}{c}{Stddev} \\
Metric	&	E	& R&	E	& R&	E	& R&	E	& R \\
\hline
Stars		& 15	& 0	& 80	& 51	& 355	& 11	& 1661	& 221 \\
Commits		& 30	& 0	& 304	& 383	& 1159	& 70	& 5323	& 1196 \\
PRs		& 0	& 0	& 25	& 42	& 161	& 3	& 672	& 94 \\
Authors		& 1	& 0	& 26	& 5	& 27	& 2	& 213	& 10 \\
Committers	& 1	& 0	& 26	& 5	& 22	& 2	& 208	& 7 \\
\hline
\end{tabular}
\vspace{-3ex}
\end{table}

We compared the earlier version of this dataset mentioned in
Section~\ref{sec:evaluation} against the Reaper dataset
of engineered software projects~\cite{MKCN17} in terms of
stars, commits, pull requests (PRs), authors, and committers
(see Table~\ref{tab:metrics}).
% select count(*) from reaper.reaper_projects;
Reaper initially contained 1\,853\,205 projects in the form \emph{login-name/project-name},
% select count(*) from reaper.reaper_ids;
from which 1\,849\,500 were successfully associated
with a project ID of GHTorrent.
Null values were substituted with zero in both datasets,
thus metrics were calculated on the basis of the entire dataset sizes
(17\,252 for this, 1\,849\,500 for the Reaper).
It appears that in all dimensions this dataset is considerably richer
than the Reaper one.
The difference most likely stems from this dataset's considerable selectivity,
as it contains two orders of magnitude fewer projects than Reaper.

\section{Related Work} % {{{1
\label{sec:related}
While the relationship between academic or semi-academic institutions
and open source software has been favorable~\cite{LT01},
with large open source projects
such as the Berkeley Software Distribution (BSD)~\cite{Ray99}
originating from them,
this has not always been the case for business.
The relationship between business
and open source software was often tense in the past,
with GPL-licensed software described as
``an intellectual property destroyer'',
un-American, and
``a cancer''~\cite{Mil02}.
Meanwhile,
others asserted that open source was compatible
with business~\cite{Hec99},
and researchers quickly identified several business models
that are based on open source software~\cite{BGR06,ASKG10},
as well as significant industrial adoption
of open source software products~\cite{SG11}.
In short,
research associated with the involvement of enterprises
in open source software can be divided
into four areas~\cite{HO11}:
a) company participation in open source development
communities~\cite{BLMR07,Hen08};
b) business models with open source in commercial
organizations~\cite{BGR06,HK10};
c) open source as part of component based software
engineering~\cite{AHCF09,LCBT09}; and
d) usage of open source processes within a company~\cite{LRM08,GFS09}.

We consider our study part of the first area.
According to Bonaccorsi et al.~\cite{BLMR07},
companies participated in one third
of the most active projects on SourceForge as
project coordinators,
collaborators in code development, or
code providers.
Hauge et al.~\cite{HSR07} also identified the role
of component integrator.
By providing their proprietary software to the open source community,
companies can benefit from
reduced development costs,
advanced performance,
repositioning in the market, and
additional profit from new services~\cite{HO11}.
Still,
the provided software should be accompanied
by adequate documentation and information
to help the community members engage in it~\cite{HSR07}.

Although companies marginally participated
in open source projects in the past,
the participation has recently increased,
especially in the larger and more active projects,
with a crucial part of the open source code being provided
by commercial organizations,
particularly small and medium-sized enterprises (SMEs)~\cite{LLL06}.
For instance,
6\%--7\% of the code in the Debian GNU/Linux distribution
over the period 1998--2004 was contributed by corporations~\cite{RDG07}.

\ifx\extended\relax
Bird et al. in their study regarding email social networks~\cite{BGDG06}
faced the challenge of duplicate email aliases
while matching identities of the email archives of Apache
to identities of Concurrent Version System (CVS) repositories.
The issue was resolved
by extracting email headers that included the sender
to produce a list of $<name, email>$ identifiers.
The similarity of the identifier pairs of the list
was then computed through a clustering algorithm,
and the resulting clusters were manually evaluated.
\fi

Similarly,
German and Mockus~\cite{GM03} linked identical contributors
of CVS repositories with multiple names or emails
of different spelling.
Using their infrastructure they identified the top contributors
of the Ximian Evolution project,
and found that the top ten contributors were Ximian employees
and consultants,
and also that private companies
such as RedHat, Ximian and Eazel,
severely affected the development of the GNOME project~\cite{Ger02},
similarly to the way the Mozilla project was mainly developed
by Netscape employees~\cite{MFH02}.

\section{Research Ideas} % {{{1
\label{sec:ideas}
The provided dataset can be employed in various ways.
First, it can be used to study
the involvement of enterprises in OSS development
by examining whether they are mostly \emph{takers} or \emph{givers},
their roles within projects, and
how they shape a project's evolution and success~\cite{BLMR07}.
Second, it can be employed in studies regarding
OSS business models,
to investigate how their choice is affected
by different enterprise characteristics such as
the employees' education level,
the enterprise's age, size, service variety,
and whether it is family-owned or not~\cite{HK10}.
Third,
it can be used for research on
the composition and structure of OSS supply chains and value chains,
particularly to identify
the added, deleted, and unchanged dependencies 
and their effect between releases
for different types of packages
such as build and test tools~\cite{DM18}.
Furthermore,
it can be employed in studies concerning
enterprise-driven global software development,
to measure benefits and tackle issues induced
from the physical separation among project members such as
strategic, cultural, communication, and knowledge management issues~\cite{HD01}.
Another use involves
identifying product or process differences between enterprise
and volunteer-driven software development in terms of
cost,
service and support,
innovation,
security,
usability,
standards,
availability,
transparency, and
reliability~\cite{PM13}.
Finally, it can be used to study
enterprise regulatory, compliance, and supply chain risks,
to investigate the risk domains
that enterprises face when engaging in OSS development,
the available sources of risk mitigation,
and the heuristics by which managers apply this understanding
to manage such projects.
From these insights,
formalized risk mitigation instruments
and project management processes can be developed~\cite{GYMK12}.

\ifx\extended\relax
\subsection*{Acknowledgements}
\else
\begin{acks}
\fi
This work has received funding from the
European Union's Horizon 2020 research
and innovation programme under grant
agreement No. 825328.

\ifx\extended\relax
D. Spinellis created the dataset and its unit tests.
Z. Kotti evaluated the rating results and researched related work.
Both contributed equally to the paper's writing.
K. Kravvaritis and G. Theodorou rated the dataset's repositories.
P. Louridas implemented the bootstrap method.
\else
\end{acks}
\fi

\bibliographystyle{ACM-Reference-Format}
\ifx\mypubs\relax
\bibliography{macro,IEEEabrv,ddspubs,myart,classics,coderead,unix,various,mybooks,bigdata,ieeestd,isostd}
\else
\bibliography{indoss,dds}
\fi % mypubs

\ifx\extended\relax
\appendix
\section{Appendix: Key SQL Queries}
% See https://tex.stackexchange.com/a/78020/10140
\pdfsuppresswarningpagegroup=1
\input{appendix}
\fi

\end{document}